\documentclass[pra,twocolumn,superscriptaddress]{revtex4}
\usepackage{color}

\usepackage{amsfonts}
\usepackage{amsmath}
\usepackage{amssymb}
\usepackage{graphicx}

\begin{document}

\title{Y-junction of Luttinger-liquid wires out of equilibrium}
\author{D.~N. Aristov}
\affiliation{NRC ``Kurchatov Institute", Petersburg Nuclear Physics Institute, Gatchina
188300, Russia}
\affiliation{Institute for Nanotechnology, Karlsruhe Institute of Technology, 76021
Karlsruhe, Germany }
\affiliation{St.Petersburg State University, 7/9 Universitetskaya nab., 199034 St.~Petersburg, Russia}
\author{I.~V. Gornyi}
\affiliation{Institute for Nanotechnology, Karlsruhe Institute of Technology, 76021
Karlsruhe, Germany }
\affiliation{A.F. Ioffe Physico-Technical Institute, 194021 St.~Petersburg, Russia}
\author{D.~G. Polyakov}
\affiliation{Institute for Nanotechnology, Karlsruhe Institute of Technology, 76021
Karlsruhe, Germany }
\author{P. W\"olfle}
\affiliation{Institute for Nanotechnology, Karlsruhe Institute of Technology, 76021
Karlsruhe, Germany }
\affiliation{Institute for Condensed Matter Theory, Karlsruhe Institute of Technology,
76128 Karlsruhe, Germany}

\begin{abstract}
We calculate the conductances of a three-way junction of spinless Luttinger-liquid wires as functions of bias voltages applied to three independent Fermi-liquid reservoirs. In particular, we consider the setup that is characteristic of a tunneling experiment, in which the strength of electron-electron interactions in one of the arms of the junction (``tunneling tip") is different from that in the other two arms (which together form a ``main wire"). The scaling dependence of the two
independent conductances on bias voltages is determined within a fermionic renormalization-group approach in the limit of weak interactions. The solution shows that, in general, the conductances scale with the bias voltages in an essentially different way compared to their scaling with the temperature $T$. Specifically, unlike in the two-terminal setup, the nonlinear conductances cannot be generically obtained from the linear ones by simply replacing $T$ with the ``corresponding''
bias voltage or the largest one. Remarkably, a finite tunneling bias voltage prevents the interaction-induced breakup of the main wire into two disconnected pieces in the limit of zero $T$ and a zero source-drain voltage.
\end{abstract}

\maketitle

\section{Introduction}
\label{intro}

Models of one-dimensional quantum wires in the presence of electron-electron interactions have been intensively studied since the 1950s \cite{Tomonaga1950,Luttinger1963}, mostly in the framework of the Tomonaga-Luttinger liquid model (TLL) featuring a linear electron dispersion relation \cite{giamarchi04}. Of particular interest is charge
transport through junctions of TLL wires. It is well recognized that the charge screening at a two-lead junction, in the limit of zero temperature $T$, may block transport completely or may enable ideal transport, depending on whether the electron-electron interaction is repulsive or attractive \cite{Kane1992}. The $T$ dependence of the conductance obeys, then, power-law scaling with an exponent that depends on the strength of interactions in the TLL wires. It is also understood that the nonlinear conductance, at a finite bias voltage $V$, in the limit of vanishing $T$ is described by a power law in $V$ whose exponent is identical to that for the $T$ dependence in the linear response \cite{Safi1995,Furusaki1996,Sassetti1996,Egger2000, Dolcini2003,Dolcini2005,Metzner2012}. This implies that the scaling behavior of the nonequilibrium conductance with varying $V$ may simply be obtained from the linear conductance by replacing $T$ with $V$.

In the present paper, we focus on a three-lead junction (``Y-junction") to which two bias voltages $V_{a}$ and $V_{b}$ 
can be applied, and derive the scaling behavior of the conductances as functions of $V_{a,b}$. 
One of the most relevant experimental setups that correspond to the three-way junction consists of a tunneling 
tip contacting a quantum wire (``main wire" below), in which case $V_a$ and $V_b$ are the source-drain and tunneling bias voltages, respectively.
We show that neither of the ``simple'' prescriptions---replacing $T$ in the linear conductance with the corresponding 
voltage ($V_b$ for the tunneling conductance and
$V_a$ for the conductance of the main wire) or with the largest one for each of the conductances---is applicable 
in the case of multi-lead junctions, where there are two or more bias voltages.
Moreover, generically, the scaling behavior of the conductances with the bias voltages is essentially different compared to their scaling with $T$.
It is worth mentioning that the possibility of an intricate interplay between different cutoffs for the scaling of
currents and noise in a nonequiulibrium Y-junction was pointed out in Ref. \cite{safi09}, where the question of whether 
the tunneling bias stops the renormalization of the junction was raised.

We build on the works on linear transport through Y-junctions \cite{Aristov2010,Aristov2011,Aristov2012,Aristov2013} 
and nonlinear transport through two-lead junctions \cite{Aristov2014} that provide a general framework for understanding
the nonequilibrium properties of multi-lead systems.
We use a fully fermionic formalism \cite{Yue1994} which explicitly assumes the existence
of thermal Fermi leads and thus avoids, by construction, complications arising within
the conventional bosonization approach when interactions are only present inside a finite segment
of a wire \cite{Maslov1995,Oshikawa2006,fnote1}. The perturbative fermionic RG theory,
formulated in Ref. \cite{Yue1994}, was efficiently used for the problem of
a double barrier \cite{nazarov03,Polyakov2003} and of a Y-junction \cite{Lal2002,Aristov2010,shi16}
in TLL. The theory has been extended to an arbitrary interaction strength by summing 
up an infinite series in perturbation theory (ladder summation).
The results for the beta-function to two-loop order and for the scaling exponents at the fixed points 
within this approach are in complete agreement with known exact results 
\cite{Aristov2008,Aristov2009,Aristov2011a,Aristov2012}. This theory was further developed to
calculate the conductance for a nonequilibrium two-lead junction (biased by a finite 
source-drain voltage) \cite{Aristov2014}.

\section{Conductances of a Y-junction}

We start by identifying the scaling terms in the perturbative expansion of the conductances to first order in the interaction strength. These are the terms that diverge logarithmically as $\ln\epsilon$ with decreasing infrared cutoff in energy space $\epsilon$. Here, we do not attempt to derive a loop expansion for the renormalization group (RG) at nonequilibrium from an effective low-energy model. Rather, we rely on the perturbative RG formalism which assumes that scaling for a given number of the coupling constants (two in our case) holds at higher orders in their perturbative expansion in the interaction strength, i.e., that the second-order terms in the expansion that diverge as $\ln^2\epsilon$ are all generated by the RG equations, etc. 

Moreover, to demonstrate the peculiar scaling behavior of the conductances at nonequilibrium, we restrict ourselves here to the case of the lowest-order perturbative RG, i.e., only keep the terms in the beta-functions (for two conductances) of the first order in the interaction strength. This corresponds to the one-loop RG taken to the limit of weak interaction. Already at this level the similarity between the scalings with the bias voltages and $T$ proves to be broken and the fixed points show a nontrivial dependence of the conductances on $V_a$ and $V_b$. The renormalizability of the nonequilibirum Y-junction model to one-loop order will be addressed elsewhere \cite{aristov17}.

\begin{figure}[tbp]
\includegraphics[width=0.85\columnwidth]{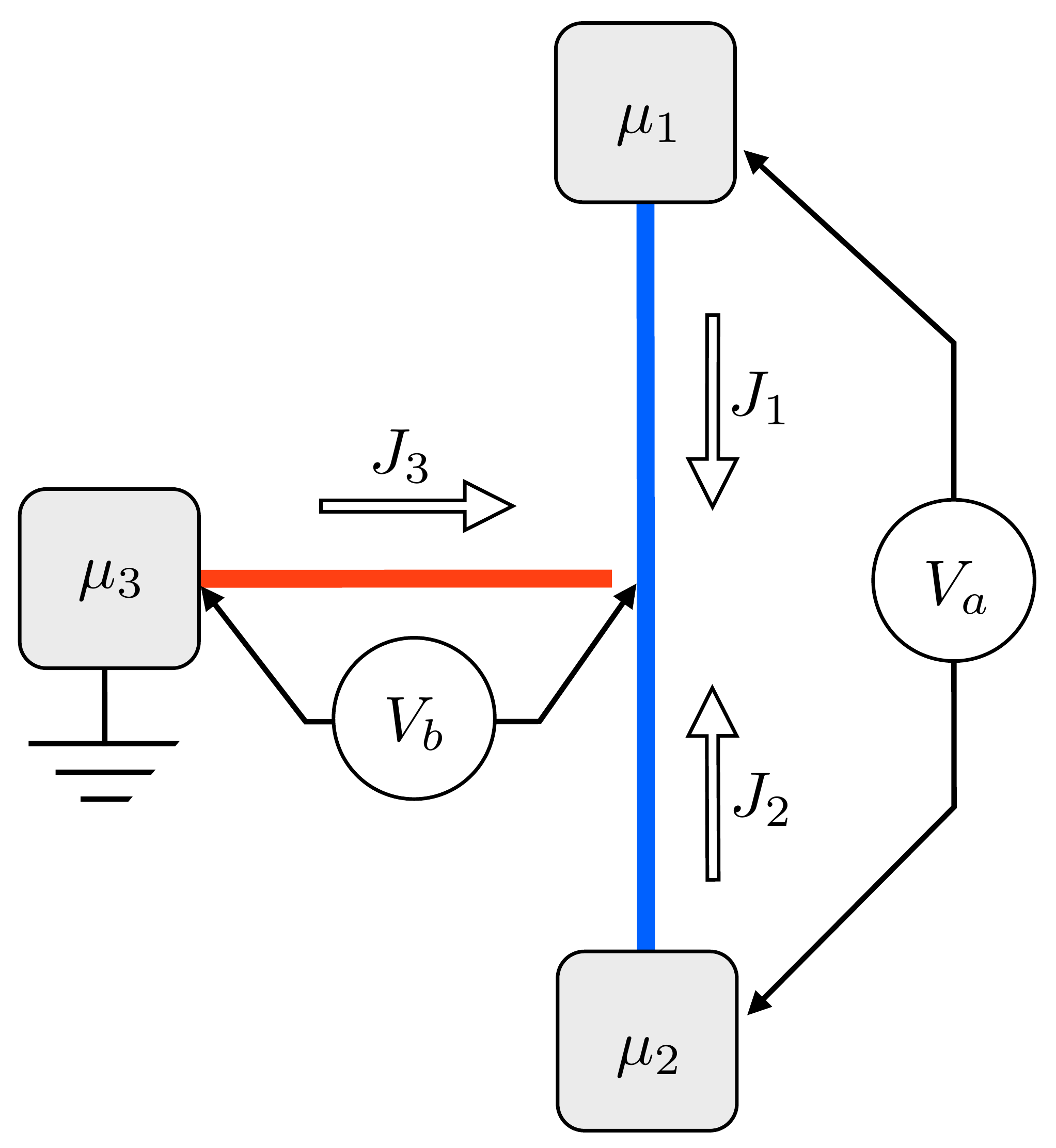}
\caption{\label{fig:setup}
Setup of the Y-junction out of equilibrium.  
The main wire is shown as a blue vertical line, the tunneling tip as a red horizontal line. The reservoirs at the chemical potentials $\mu_{a,b,c}$ are depicted as gray blocks, with the currents $J_{a,b,c}$ flowing out from them in the presence of the bias voltages $V_a$ and $V_b$. 
}
\end{figure}

We consider a  Y-junction whose main wire is characterized by the dimensionless interaction
constant $\alpha$ across its whole length. The junction is thus symmetric in the sense that 
the interaction constants $\alpha_1$ and $\alpha_2$ in arms $1$ and $2$, respectively, 
are equal to each other, $\alpha_1=\alpha_2\equiv \alpha$. The main wire is
connected to reservoirs with chemical potentials $\mu _{1}$ and 
$\mu_{2}$, as shown in Fig.\ \ref{fig:setup}. The third arm of the junction is a tunneling-tip wire, with the interaction constant 
$\alpha_{3}$, connected to a reservoir at a chemical potential $\mu _{3}=0$. 
The currents $J_{i}$  flowing from the reservoirs $i=1,2,3$ with the chemical potentials $\mu_i$ 
towards the junction obey Kirchhof's law, $\sum_{i}J_{i}=0$. 

It is convenient to introduce two independent currents $J_{a,b}$ and two independent bias voltages $V_{a,b}$ as follows:
\begin{equation}
J_{a}=\frac{1}{2}(J_{1}-J_{2}), \quad V_{a}=\mu _{1}-\mu _{2}
\end{equation}
for the main wire and
\begin{equation}
J_{b}=\frac{1}{3}(J_{1}+J_{2}-2J_{3})=-J_{3}, \quad V_{b}=\frac{1}{2}(\mu_{1}+\mu _{2}) 
\end{equation}
for the tunneling tip. The conductances are then defined as 
\begin{equation}
\begin{aligned}
J_{a}&=G_{a}V_{a}+G_{ab}V_{b}, \\
J_{b}&=G_{b}V_{b}+G_{ba}V_{a}. 
\end{aligned}
\label{J}
\end{equation}
For the main wire which is symmetric under interchange of the ends, the off-diagonal
conductances vanish in the linear-response limit, 
\begin{equation}
G_{ab}(V_{a}\rightarrow 0,V_{b}\rightarrow 0)=G_{ba}(V_{a}\rightarrow 0,V_{b}\rightarrow 0)=0. 
\end{equation}
This is no longer true in the nonequilibrium case, where $G_{ab}\neq G_{ba}$, 
except if $|\mu _{1}|=|\mu _{2}|$, meaning that the system is symmetric 
with respect to interchanging the two ends of the main wire, including the chemical 
potentials. The reason why the case $\mu_{1}=-\mu_{2}$ is also referred to as ``symmetric" 
is the presence of the time-reversal invariance which implies that the cases 
$\mu _{1}=\mu $, $\mu_{2}=-\mu $ and $\mu _{1}=-\mu $, $\mu _{2}=\mu $ are completely equivalent.
We will see, though, that $G_{ab}$, $G_{ba}$ can be expressed in terms of the diagonal conductances.

We formulate our model in the scattering-state basis.
The incoming ($\psi_{j}^{\text{in}}$) and outgoing ($\psi_{j}^{\text{out}}$) electronic waves
at the junction ($x\to 0$) are related  by the $S$ matrix 
$\Psi_\text{out}(x)=S\cdot \Psi_\text{in}(x)$, 
where  
$\Psi_\text{in}=(\psi _{1}^{\text{in}},\psi _{2}^{\text{in}},\psi _{3}^{\text{in}})^\text{T}$
and similar for $\Psi_\text{out}$.
The scattering matrix $S$ characterizing the symmetric junction 
is given explicitly in  Appendix \ref{app:deriv1}.
The unitarity of the $S$-matrix imposes the constraint 
\begin{equation}
0\leq G_{b}\leq 1-4\left(G_{a}-\frac{1}{2}\right)^{2},
\label{bound}
\end{equation}
such that the region of allowed
conductance values in the $G_{b}$-$G_{a}$ plane is bounded by a parabola
\cite{Aristov2010}.

The Hamiltonian of the TLL model reads, then, as: 
\begin{equation}
\begin{aligned}
\mathcal{H} &=\int_0^{\infty}\!\!dx\,[H_{0}+H_\text{int}\, \theta(x-\ell)\theta(L-x)]\,,   \\
H_{0} &= i v \left(
\Psi_\text{in}^{\dagger }\nabla \Psi _\text{in}-\Psi_\text{out}^{\dagger }\nabla \Psi_\text{out}\right)\,, \\
H_\text{int} &= 2\pi v  \sum\limits_{j=1}^3  \alpha_{j}
 \widehat{\rho }_{j} \widehat{\widetilde{\rho }}_{j} \,,
\end{aligned}
\label{Ham}
\end{equation}
with the interaction present only in the interval $\ell<x<L$.
The incoming ($\widehat{\rho }_{j}$) and outgoing ($\widehat{\widetilde{\rho }}_{j}$) density operators are defined by 
\begin{eqnarray}
\widehat{\rho }_{j}=\Psi ^{+}_\text{in}\rho _{j}\Psi_\text{in}, \quad
\widehat{\widetilde{\rho }}_{j}=\Psi ^{+}_\text{in}\widetilde{\rho }_{j}\Psi_\text{in},
\end{eqnarray} 
where $\widetilde{\rho }_{j}=S^{+}\cdot \rho _{j}\cdot S$
and the matrix $\rho _{j}$ is given by $(\rho _{j})_{\alpha \beta }=\delta _{\alpha \beta }\delta _{\alpha j}$, so that 
$(\widetilde{\rho }_{j})_{\alpha \beta }=S_{\alpha j}^{+}S_{j\beta }$. 
The short-distance scale $\ell$ defines the ultraviolet energy cutoff $\omega_{0} = v/\ell$; we set $v=1$ below.

\begin{figure}[tbp]
\includegraphics[width=0.95\columnwidth]{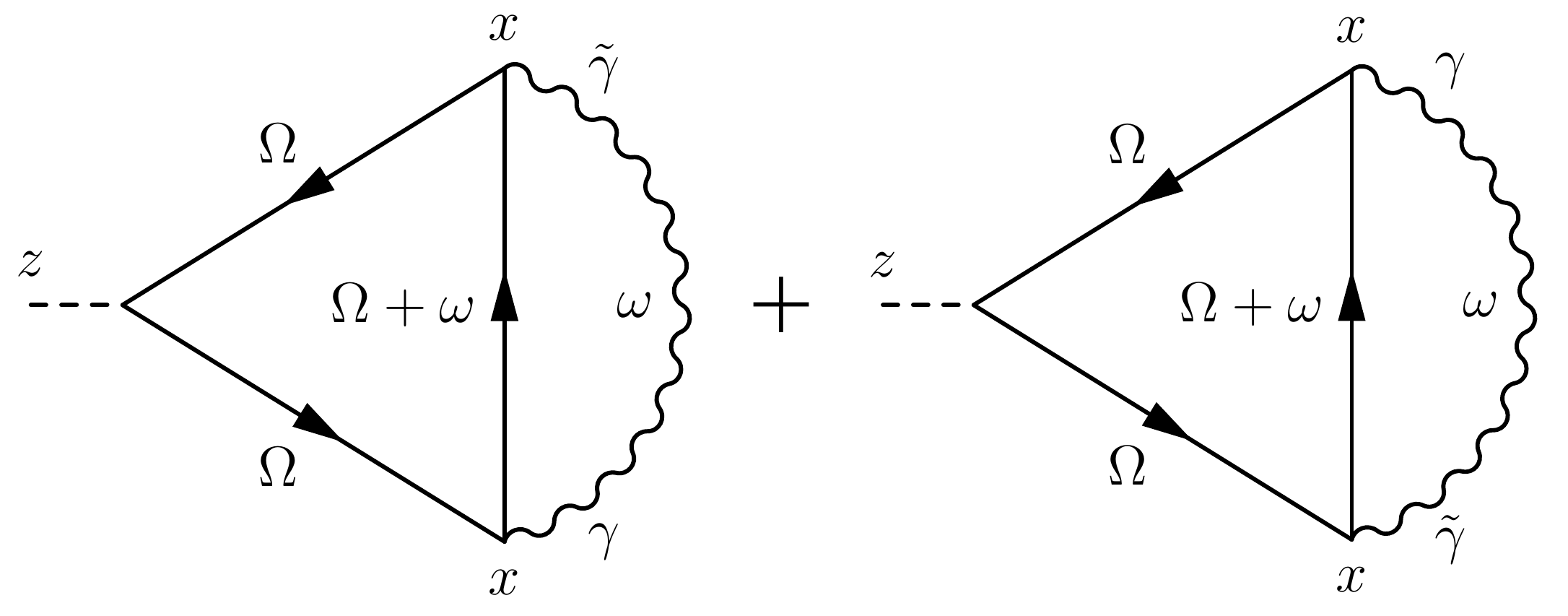}
\caption{\label{fig:1order}
Diagrams for the corrections to the currents to first order in interactions.
The current is calculated at point $z$ and the interaction takes place at point $x$
(both $z$ and $x$ include the space coordinate and the wire index).
The solid and wavy lines stand for the electron Green's functions and interaction, respectively.
The fermion-boson vertices in Keldysh space are denoted by the matrices $\gamma$ and $\tilde{\gamma}$;
for more detail, see Appendix~\ref{app:deriv1} and Ref.~\cite{Aristov2014}.
}
\end{figure}

We consider the currents as functions of the infrared cutoff $\epsilon =\omega _{0}e^{-\Lambda }$.
The first-order interaction-induced corrections to the currents are described by 
the diagrams in Fig.~\ref{fig:1order}. The calculation, similar to that in Ref.~\cite{Aristov2014}, 
is presented for the Y-junction geometry in Appendix \ref{app:deriv1}.
Without interaction, the currents are given by 
$J_{a}^{(0)} = G_{a}^{(0)} V_{a}$, 
$J_{b}^{(0)} = G_{b}^{(0)} V_{b}$. 
The scale-dependent corrections to the currents at
$T=0$ in the limit $L\rightarrow \infty $, are written, to first order in the interaction, as:
\begin{equation}
\begin{aligned} 
J_{a,b}^{(1)}&=\int_{\epsilon }^{\omega _{0}}\!\!d\omega\,
R_{a,b}(\omega)\,,
\\ R_{a}& =A_{1}F(\omega,V_{a})+ A_{2}[F(\omega,V_{b+})-F(\omega ,V_{b-})]\,,
\\ R_{b}& =B_{2}[F(\omega,V_{b+})+F(\omega ,V_{b-})]\,,
\end{aligned}
\label{corrections}
\end{equation}%
where we defined 
$$F(\omega ,V)=(|\omega + V|-|\omega -V|)/\omega$$ 
and 
$$V_{b_\pm}=V_{b}\pm V_{a}/2.$$ 
We choose,
without a loss of generality, $V_{a,b} \geq 0$, so that $|V_{b-}| \leq V_{b+}$.
The prefactors $A_{1},A_{2},B_{2}$ are found in the form
\begin{equation}
\begin{aligned} A_{1} & =-\alpha\left[G_{a}(1-G_{a})-\frac{G_{b}}{4}\right] , \\ A_{2} &
=-\frac{\alpha+(\alpha+2\alpha_{3})(1-2G_{a})}{16}G_{b} , \\ B_{2} &
=-\frac{\alpha(1-2G_{a})+(\alpha+2\alpha_{3})(1-G_{b})}{8}G_{b} . \end{aligned}
\end{equation}
We note that if the tip is detached from the main wire, so that $G_{b}=0$, we
recover the equations for a two-lead wire from Ref.~\cite{Aristov2014}.

The $\omega$ integration of $F(\omega ,V)$ yields
\begin{align} 
&\mathcal{I}(\omega _{0},\epsilon ,V)
 =\int_{\epsilon }^{\omega_{0}}\!d\omega \, F(\omega ,V)  
  =\theta (\epsilon -|V|)\,2V\ln \frac{\omega_{0}}{\epsilon }\notag \\ 
 &+2\theta (|V|-\epsilon )\left[V-\epsilon \,\text{sgn}(V)+V\ln
\frac{\omega _{0}}{V}\right] , \notag \\ 
& \simeq \theta (\epsilon -|V|)\, 2V\ln \frac{\omega _{0}}{\epsilon } \, .
\label{I0}
\end{align}
In the last line of Eq.~(\ref{I0}), we neglected the contribution of $\epsilon<|V|$.
This is because it is not an infrared-divergent scaling contribution, being only linearly
dependent on $\epsilon $ (so that its derivative with respect to 
$\Lambda$ vanishes exponentially in $\Lambda \rightarrow \infty$). 
We thus obtain the scaling contributions to the currents as 
\begin{align} 
J_{a}(\epsilon )& =J_{a}(\omega_{0})+2\left[A_{1}V_{a}\, \theta (\epsilon-V_{a})+A_{2}V_{b+}\, \theta (\epsilon -V_{b+})
\right. \notag \\
& \left. -A_{2} V_{b-}\,
\theta (\epsilon -|V_{b-}|)\right] \ln \frac{\omega _{0}}{\epsilon }\, ,
\end{align}
and
\begin{equation}
\begin{aligned} J_{b}(\epsilon )& =J_{b}(\omega _{0})+2B_{2}[V_{b+}\,
\theta (\epsilon-V_{b+}) \\ & +V_{b-}\,\theta (\epsilon -|V_{b-}|)]
\ln\frac{\omega _{0}}{\epsilon }\,,
\end{aligned}
\end{equation}
where we set $J_{a}^{(0)}=J_{a}(\omega _{0})$, etc.
An extension of this calculation to the case of finite $T$ is presented in Appendix~\ref{app:T}.

The corrections to the conductances are then given by
\begin{equation}
\begin{aligned} 
G_{a}(\epsilon )& =G_{a}(\omega _{0})
+2 A_{1} \Lambda\, \theta (\epsilon-V_{a})+ A_{2} \Lambda \, \theta _{+}(\epsilon )\,,
\\ G_{b}(\epsilon )&
=G_{b}(\omega _{0})+2B_{2}\, \Lambda\, \theta _{+}(\epsilon )
 \,,
\end{aligned}
\label{14}
\end{equation}
and
\begin{equation}
\begin{aligned} 
G_{ab}(\epsilon )& =-2A_{2} \Lambda \, \theta _{-}(\epsilon )\,,
\\ 
G_{ba}(\epsilon )&
=-B_{2}\Lambda \,
\theta _{-}(\epsilon)\,,
\end{aligned}
\label{15}
\end{equation}%
where 
$$\theta _{\pm }(\epsilon )
=\theta (\epsilon - |V_{b-}|)\pm \theta (\epsilon -V_{b+})$$
takes the value of 2 for $\epsilon>V_{b+}$ and 0 for $\epsilon<|V_{b-}|$, and 1 in between.
The off-diagonal conductances are only nonzero in the energy window $|V_{b-}|<\epsilon <V_{b+}$.
It is worth noting that, since $A_{2}$ and $B_{2}$ depend only on the
diagonal conductances, $G_{ab}$ and $G_{ba}$ are not independent quantities.

\section{RG equations to first order in the interaction}

We now assume that the conductances obey scaling behavior (to be verified in Ref.~\cite{aristov17}), 
as expressed by
\begin{equation}
G_{a,b}(\epsilon;V_{a},V_{b},\omega _{0})=
\Phi_{a,b}\left(\frac{\epsilon}{\omega _{0}},\frac{V_{a}}{\omega _{0}},\frac{V_{b}}{\omega _{0}}\right) \,,
\end{equation}
where $\Phi_{a,b}(x,y,z)$ is the scaling function.
Differentiating this relation with respect to $\Lambda$, with the use 
of Eqs.~(\ref{14}) and (\ref{15})
we get the perturbatrive RG equations
\begin{equation}
\begin{aligned} 
\frac{dG_{a}}{d\Lambda }& \equiv\beta_{a}(G_{a},G_{b})
=2A_{1}\theta (\epsilon -V_{a})+A_{2}\theta _{+}(\epsilon ) \,,
\\ 
\frac{dG_{b}}{d\Lambda }& \equiv\beta _{b}(G_{a},G_{b})=2B_{2}\theta
_{+}(\epsilon ) \,,
\end{aligned}
\label{RG}
\end{equation}%
and
\begin{equation}
\begin{aligned} \frac{dG_{ab}}{d\Lambda }& =-2A_{2}\theta _{-}(\epsilon ) \,,  \\
\frac{dG_{ba}}{d\Lambda }& =-B_{2}\theta _{-}(\epsilon )   \,,
\end{aligned}
\label{RG1}
\end{equation}%
with the initial values $G_{a,b}(\omega _{0})$. 
The three quantities 
$A_{1},A_{2},B_{2}$ 
will be taken to depend on the flowing conductances
determined by the RG method. Since they do not depend on
the off-diagonal conductances, the RG equations Eq.~\eqref{RG} form a closed
system determining the diagonal conductances. 

It is important to emphasize that the dependence of the right-hand side
of the RG equations \eqref{RG} on $\epsilon$ through the step functions
only means that the renormalization occurs in several steps with different 
beta-functions at each step. 
Specifically, Eqs.~\eqref{RG} and \eqref{RG1} lead to different scaling behavior of the conductances 
in three distinct cases: 
\begin{eqnarray}
V_{a}<\tfrac{2}{3}V_{b},\quad \tfrac{2}{3}V_{b}<V_{a}<2V_{b},\quad 2V_{b}<V_{a}.
\label{3cases}
\end{eqnarray}
At high energies, $\epsilon > \max\{V_{a},V_{b+}\}$, the
scaling follows the behavior found in the linear-response regime
\cite{Lal2002,Aristov2010} in all these cases. 
On the other hand, below $\max\{V_{a},V_{b+}\},$
the scaling depends on which of the cases specified in Eq.~(\ref{3cases}) is realized.

It follows from the expressions for $A_{1},A_{2},B_{2}$ in terms of $%
G_{a},G_{b}$ that in the limit $V_{a,b}\to 0$, the system of RG equations \eqref{RG} 
possesses three fixed points, i.e., there are three solutions of the pair of
equations $\beta _{a}(G_{a},G_{b})=\beta _{b}(G_{a},G_{b})=0$ (we restrict
our discussion to repulsive interactions; see the discussion in Ref.~\cite{Aristov2011a}). 
There is one stable fixed point, termed $N$, at $G_{a}=G_{b}=0$, and there are two unstable fixed points,
$A$ at $G_{a}=1$, $G_{b}=0$ and $M$ at $G_{a}^{M}$, $G_{b}^{M}$. All the fixed points
belong to the boundary of the range of allowed conductances in the $G_{b}$-$G_{a}$ plane,
specified above. Below, we discuss scaling of the conductances in the vicinity of 
the stable fixed point. The analysis of the nonlinear conductances near the unstable
fixed points is given in Appendix~\ref{app:unstable}.

\section{Stable fixed point}

Near the fixed point $N$, we linearize the 
beta-function in $G_{a},G_{b}$. By introducing the combination 
$$G_{c}=G_{a}-\frac{1}{4}G_{b},$$ 
we eliminate the terms in the RG equation
for $G_{c}$ that are proportional to $\theta _{+}(\epsilon )$ and get
 \begin{equation}
\begin{aligned}
\frac{dG_{c}}{d\Lambda }& =-2\alpha G_{c}\theta (\epsilon -V_{a}) \,, \\
\frac{dG_{b}}{d\Lambda }& =-\frac{\alpha+\alpha_{3}}{2}G_{b}\theta _{+}(\epsilon )  \,.
\end{aligned}
\end{equation}
The solution for $G_{c}$ reads:
 \begin{equation}
\begin{aligned}
\frac{G_{c}(\epsilon )}{G_{c}(\omega _{0})}& =
\left\{
  \begin{array}{ll}
    \left(\epsilon/\omega _{0}\right) ^{2\alpha}, & \epsilon >V_{a}, \\[0.2cm]
    \left( V_{a}/\omega _{0}\right) ^{2\alpha}, & \epsilon <V_{a},
  \end{array}
\right.\\
\end{aligned}
\end{equation}
where
\begin{equation}
G_{c}(\omega _{0})=G_{a}(\omega _{0})-\tfrac{1}{4}G_{b}(\omega _{0}) \,.
\end{equation}

Recall that we have assumed positive voltages $V_{a,b}>0$, such that $V_{b+}>|V_{b-}|$. 
For large $\epsilon >V_{b+}$, we find
\begin{equation}
G_{b}(\epsilon )=G_{b}(\omega _{0})\left( \frac{\epsilon }{\omega _{0}}%
\right) ^{\alpha+\alpha_{3}}\,, \quad\epsilon >V_{b+} \,.
\end{equation}%
In the intermediate range $|V_{b-}|<\epsilon <V_{b+}$, the scaling exponent is reduced by a factor of 2:
\begin{equation}
G_{b}(\epsilon)=G_{b}(V_{b+})\left( \frac{\epsilon }{V_{b+}}\right)^{(\alpha+\alpha_{3})/2}\,,
\quad |V_{b-}|<\epsilon <V_{b+} \,.
\end{equation}%
For the lowest running energies, $\epsilon <|V_{b-}|$, the scaling stops and 
$$G_{b}(\epsilon)=G_{b}(|V_{b-}|)\,,\quad \epsilon <|V_{b-}|.$$

Summarizing, the diagonal conductances near the $N$ point are found as
\begin{eqnarray}
G_{b}(0) &=&G_{b}(\omega _{0})
\left( \frac{\sqrt{|V_{b}^{2}-V_{a}^{2}/4|}}{\omega_{0}}\right)^{\alpha+\alpha_{3}}\,,  \label{Gbpowerlaw} \\
G_{a}(0) &=&\left[G_{a}(\omega _{0})-\frac{1}{4}G_{b}(\omega _{0}) \right]
\left( \frac{V_{a}}{\omega _{0}} \right) ^{2\alpha}+\frac{1}{4}G_{b}(0)
\,.\nonumber 
\\  \label{Gapowerlaw}
\end{eqnarray}%
Note that the renormalized conductance $G_{b}(0)$ exhibits a split zero-bias anomaly
(i.e., vanishes to zero) at $V_b=\pm V_a/2$ (or, equivalently, $\mu_1=\mu_3$ or $\mu_2=\mu_3$). 
This is similar to the splitting of the
tunneling density of states in a nonequilibrium TLL wire with a double-step
distribution function supplied by the leads \cite{gutman08}.
In our case, however, the effect of a double-step distribution is produced by scattering
off the junction. Remarkably, it is sufficient to align the chemical potential in the tunneling tip
with the chemical potential only in one arm of the main wire to make $G_b$ vanish.
One more difference is that the critical exponent for the split
zero-bias anomaly in our case (fixed point $N$) is linear in interactions, whereas in the case
of noninvasive tunneling (fixed point $A$, see Appendix~\ref{app:unstable},  with $\alpha_3=0$) it is quadratic. 
It is also worth noticing that the tunneling zero-bias anomaly also shows up in
the conductance of the main wire, Eq.~(\ref{Gapowerlaw}).

For the off-diagonal conductances, we get
 \begin{align}
&G_{ab}(0)=G_{ba}(0)=\frac{1}{2}[G_{b}(V_{b+})-G_{b}(0)] =\frac{1}{2}G_{b}(\omega _{0})\notag \\
&\times
\left[\left( \frac{V_{b}+V_a/2}{\omega _{0}}\right)^{\alpha+\alpha_{3}}
-\left( \frac{\sqrt{|V_{b}^{2}-V_{a}^{2}/4|}}{\omega _{0}}\right)^{\alpha+\alpha_{3}} \right]  \,.
\notag\\
\end{align}
The currents are obtained by means of Eqs.~(\ref{J}) that
combine the diagonal and off-diagonal conductances.
Differentiating the currents with respect to the bias voltages
yields differential conductances 
$$\tilde{G}_{a}=\frac{\partial J_{a}}{\partial V_{a}}, 
\quad
\tilde{G}_{b}=\frac{\partial J_{b}}{\partial V_{b}},
\quad \tilde{G}_{ab}=\frac{\partial J_a}{\partial V_b},
\quad \tilde{G}_{ba}=\frac{\partial J_b}{\partial V_a}.
$$
Note that such a differentiation might potentially give rise to
divergent terms in the differential conductances. However,
the singular contributions coming from the diagonal conductances are
cancelled by the terms produced by differentiation of the off-diagonal
conductances. As a result, in the limit of weak interaction, $\alpha,\alpha_3\ll 1$,
we obtain for the renormalized conductances related to the current in the main wire:
\begin{eqnarray}
\tilde{G}_{a}(0)\simeq G_{a}(0),&& \quad \tilde{G}_{ab}(0)\simeq G_{ab}(0)
\end{eqnarray}
at arbitrary voltages, including the singular points.
At the same time, in the differential tunneling conductance
we have to keep the small (in $\alpha+\alpha_3$) correction stemming 
from the off-diagonal conductance $G_{ba}$,
\begin{equation}
\tilde{G}_{b}(0)\simeq G_{b}(0)+G_b(\omega_0)\frac{(\alpha+\alpha_3)V_a}{2V_b+V_a}\left( \frac{V_b+V_{a}/2}{\omega_{0}}\right)^{\alpha+\alpha_{3}},
\end{equation}
near the zero-bias anomaly at $V_b=V_a/2\neq 0$, where $G_{b}(0)=0$.
This means that the differential conductance $\tilde{G}_{b}$ for 
tunneling into the biased wire does not completely vanish, 
$$
\left.\tilde{G}_{b}(0)\right|_{V_b=V_a/2}\simeq \frac{\alpha+\alpha_3}{2}G_b(\omega_0)\left( \frac{V_{a}}{\omega_{0}}\right)^{\alpha+\alpha_{3}}\neq 0,
$$
in contrast to $G_b$.

We now turn to the discussion of the general result for the $N$ point, 
Eqs.~(\ref{Gbpowerlaw}) and (\ref{Gapowerlaw}), in two limiting cases.
In particular, one may wonder whether in the case that one of the bias voltages is much
larger than the other, the smaller of the voltages may be neglected.
If true, this would make it possible for the linear-response scaling to
carry over to the nonlinear regime by simply replacing $T$ with
the larger of the voltages. As we show now, this is not always the case.

\subsubsection{$V_{b}\ll V_{a}$}

Consider first the case of small bias on the tunneling tip, $V_{b}\ll V_{a}$, i.e., 
$V_{b+}\simeq |V_{b-}|\simeq V_{a}/2\gg V_{b}$. 
From Eqs.~(\ref{Gbpowerlaw}) and (\ref{Gapowerlaw}),
we have
\begin{align}
G_{b}(0)& \simeq G_{b}(\omega_{0})
\left(\frac{V_{a}}{\omega_{0}}\right)^{\alpha+\alpha_{3}}   \,,  \\
G_{a}(0)& =\left[G_{a}(\omega_{0})-\frac{1}{4}G_{b}(\omega_{0})\right]
\left(\frac{V_{a}}{\omega_{0}}\right)^{2\alpha}\nonumber \\
&+\frac{1}{4}G_{b}(\omega_{0})
\left(\frac{V_{a}}{\omega_{0}}\right)^{\alpha+\alpha_{3}}  \,.
\end{align}
These results are obtainable from the linear-response conductances as a
function of $T$, $G_{b}(T)\propto T^{\alpha+\alpha_{3}}$ and $G_{a}(T)-\tfrac{1}{4}G_{b}(T)\propto T^{2\alpha}$,
by substituting $V_a$ (the \textit{larger} of the voltages) for $T$. 
Note that, despite the deceitfully simple appearance, this is rather nontrivial.
Indeed, $G_{b}$ here is controlled by $V_{a}$, which is to say that the bias voltage $V_{b}$ driving the
current $J_{b}$ does not enter the nonlinear dependence of $G_{b}$. 
Thus, one of the naive ``general'' prescriptions mentioned in Sec.~\ref{intro},
which would imply the replacement of $T$ by the \textit{corresponding} voltage
(in the present case this would be $V_b$), fails.

\begin{figure}[tbp]
\includegraphics[width=0.9\columnwidth]{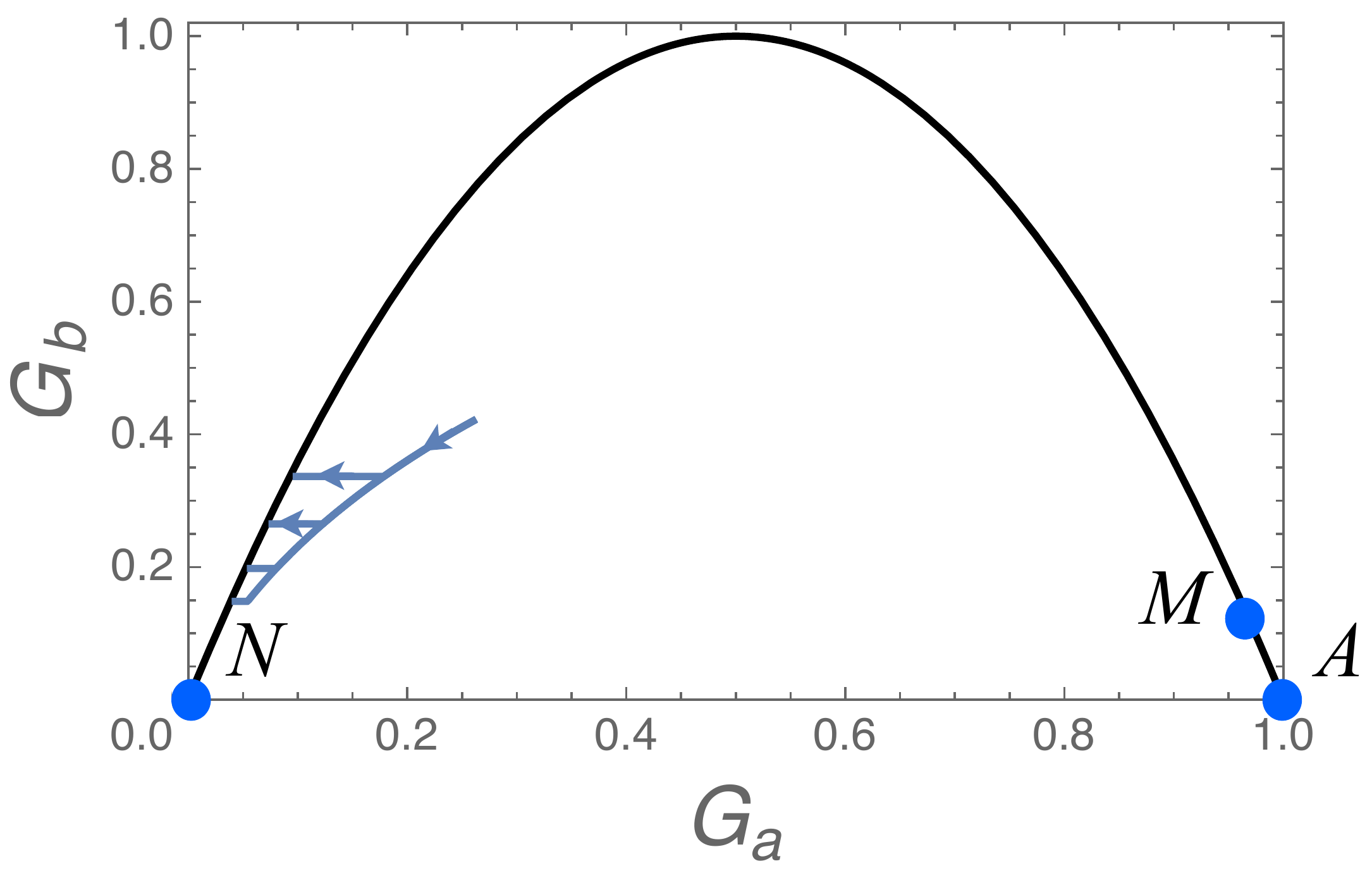}
\caption{ \label{fig:RGflows}
Zero-$T$ RG flows for different voltages $V_{a}/\omega_{0} = 0.3, 0.1, 0.03, 0.01$ (from top to bottom) for
the interaction constants $\alpha=0.3$, $\alpha_{3} =0.01$,
the tunneling bias $V_{b} = 0.3\, \omega_{0}$, and the bare conductances $G_{a}(\omega_{0})=0.26$,  $ G_{b}(\omega_{0}) = 0.42$. The position of fixed points $N$, $A$, $M$ at equilibrium is depicted by blue dots. The limiting RG values of the conductances out of equilibrium lie on the parabolic fixed line. 
}
\end{figure}

\subsubsection{$V_{b}\gg V_{a}$}

In the opposite limiting case of small bias in the main wire, $V_{b}\gg V_{a}$,
we have
\begin{align}
G_{b}(0)& \simeq G_{b}(\omega_{0})\left(\frac{V_{b}}{\omega_{0}}\right)^{\alpha+\alpha_{3}} \,,  \\
G_{a}(0)& \simeq \left[G_{a}(\omega_{0})-\frac{1}{4}G_{b}(\omega_{0})\right]
\left(\frac{V_{a}}{\omega_{0}}\right)^{2\alpha} \nonumber
\\
&+\frac{1}{4}G_{b}(\omega_{0})
\left(\frac{V_{b}}{\omega_{0}}\right)^{\alpha+\alpha_{3}} \,. \label{30}
\end{align}
Now we see that $G_{b}(0)$ is found to scale as 
$(V_{b}/\omega_{0})^{\alpha+\alpha_{3}}$, which is obtainable by replacing $T$ with $V_{b}$ in the
linear-response expression for $G_{b}(T)$, i.e., both ``naive'' prescriptions from Sec.~\ref{intro}
do work. However, $G_{a}(0)$ depends now on \textit{both} voltages, 
which cannot be obtained from the linear-response scaling of $G_a(T)$ by the replacement of $T$ with
one of the voltages. Thus, we find the remarkable result
that, depending on the initial conductances $G_{a,b}(\omega _{0})$ and the interaction constants 
$\alpha$ and $\alpha_{3}$, the conductance $G_{a}(0)$ of the main wire is
controlled by either the \textit{corresponding} voltage $V_{a}$ applied to the main
wire or by the \textit{larger} voltage $V_{b}$ applied to the tunneling tip.
We, therefore, conclude that \textit{none} of the prescriptions translating the linear and nonlinear conductances
into each other is general.

It is also interesting to look at the situation from another perspective:
by observing that the renormalization of the conductances may continue for energy
scales below the larger voltage.
In particular, the larger tunneling bias does not completely stop the RG flow for the zero-$T$ conductance
$G_a$ of the main wire. One might expect, then, that $G_a$ at $V_a\to 0$ would renormalize to zero, 
similarly to a two-lead junction. This is not the case, as seen from Eq.~(\ref{30}), which can be 
traced back to the condition (\ref{bound}) following from the unitarity of the $S$-matrix. That is, remarkably, 
a finite tunneling bias, although not stopping the RG flow for $G_a$, still prevents the
interaction-induced breakup of the main wire into two disconnected pieces, as illustrated in Fig.\ \ref{fig:RGflows}.

\section{Summary}

We have developed a framework to study nonlinear 
charge transport through a $Y$-junction, i.e., through a
junction connecting three Luttinger-liquid quantum wires.
The particular example of nonequilibrium on which we have focused in the
present paper is the setup with the tunneling tip and the main wire 
biased by the tunneling and source-drain voltages, respectively.
This setup is characterized by four nonlinear conductances connecting the two 
currents with two (source-drain, $V_a$, and tunneling, $V_b$) voltages. 
We have shown that the off-diagonal components 
of the conductance matrix are completely determined by the diagonal ones, $G_a$ and $G_b$
for the main wire and the tunneling tip, respectively. 

We have calculated the interaction-induced corrections to the currents to
first order in the interaction strength and identified the scale-dependent terms. 
This has allowed us to derive two coupled renormalization group equations for $G_a$ and $G_b$. 
We have solved these equation analytically in the vicinity of three fixed points 
known to exist in the linear-response regime. For the case of repulsive interaction, there
exists one stable fixed point at $G_{a}=G_{b}=0$, around which the
conductances obey power laws 
$G_{a}-\frac{1}{4}G_{b}\propto V_{a}^{2\alpha}$ and
$G_{b}\propto |V_{b}^{2}-V_{a}^{2}/4|^{(\alpha+\alpha_{3})/2}$
with the interaction dependent exponents. 
This is to say that neither is $G_{a}$ following a pure power law in $V_{a}$, nor
is $G_{b}$ given by a pure power law in $V_{b}$, as one might naively
expect. Remarkably, it is also seen that the RG flow is not generically 
stopped by the largest of the two voltages. Notably, the singularity related to the
split zero-bias anomaly at $V_b=\pm V_a/2$ shows up in both $G_a$ and $G_b$. 

We have thus demonstrated that, in general, the conductances scale with the bias voltages 
in a way that is essentially different from their scaling with temperature, unlike in the two-lead junctions. 
Interestingly,  even though not stopping the renormalization of $G_a$, a finite tunneling voltage $V_b$ 
precludes transport through the main wire from being blocked in the limit of zero $T$ and a zero 
source-drain voltage $V_a$. At the same time, a finite source-drain voltage $V_a$ in the main wire
precludes the differential tunneling conductance from vanishing to zero at the zero-bias anomaly.

\section{Acknowledgements}
 The work of D.A. was partly supported by RFBR grant No. 15-52-06009 and by GIF grant No. 1167-165.14/2011.

\appendix

\section{\label{app:deriv1} Derivation of the first-order corrections}

Here we describe how the first-order corrections to the nonequilibrium
currents through a Y-junction were calculated.
First of all, we define the Green's function as a quantity
dependent 
on two continuum variables, the space coordinate $x$ and energy $\omega $, 
 the Keldysh index $i=1,2$, 
 the ``in-out'' label $l=1,2$ and 
 the wire index,
which can, in general, take values $j=1,\ldots N$; in our situation $N=3$.
This is done according to Eq.~(5) in Ref.~\cite{Aristov2014}. The Keldysh
weight functions $h_{j}(\Omega )=\tanh [(\Omega -2\mu _{j})/2T]$ depend on
the chemical potential $\mu _{j}$ and the temperature $T$.

As defined in Eq.~(5) of \cite{Aristov2014}, the Green's functions depend
on the elements of the $S$-matrix, which we assume to be symmetric:
\begin{equation}
S=%
\begin{pmatrix}
r_{1}, & t_{1}, & t_{2} \\
t_{1}, & r_{1}, & t_{2} \\
t_{2}, & t_{2}, & r_{2}%
\end{pmatrix} \,.
\label{Smat}
\end{equation}%
All the Fermi velocities in the leads are assumed to be the same.

The calculated corrections to the currents are shown by the skeleton graphs in
Fig.~\ref{fig:1order}. The correction to the incoming current is zero. 
Evaluation of the correction to the outgoing current yields three types of terms, 
containing quadratic forms in the Keldysh
weights, e.g. $h_{j}(\Omega )h_{l}(\omega +\Omega)$, first powers $h_{j}(\Omega )$, $h_{j}(\omega +\Omega )$, 
and terms independent of $h_{j}$.
The integration over $\Omega$ leads to the
disappearance of the linear terms, while the terms quadratic in $h_{j}$ and those
independent of $h_{j}$ may be combined into an integrand, convergent at 
$\Omega \to \pm \infty $, of the form
\begin{equation*}
\int d\Omega [1-h_{j}(\omega +\Omega )h_{l}(\Omega )]=2f_{2}(\omega +\mu
_{l}-\mu _{j}) \,,
\end{equation*}
with    $f_{2}(x)=x\coth (x/2T)$.
The corrections to the currents $j_{a}$ and $j_{b}$ are obtained by first
integrating over $x\in \lbrack 0,L]$, which leads to the combination of the form 
$(1-e^{2i\omega L})/\omega $. At this step, we obtain intermediate expressions as
six linear combinations of $f_{2}(\omega +\mu _{l}-\mu _{j})$ and three
linear combinations of $f_{2}(\mu _{l}-\mu _{j})$ with all possible $j\neq l$.
The terms with $j=l$ are cancelled.

The last integration is over $\omega \in \lbrack -\omega _{0},\omega _{0}]$. 
Using the  property $f_{2}(x)=f_{2}(-x)$, we simplify
these expressions further and reduce the last integration to the one over the
positive interval, $\omega \in \lbrack 0,\omega _{0}]$.  The resulting corrections contain terms with
large logarithmic factors stemming from the integral [see Eq. (21) in Ref.~\cite{Aristov2014}]
\begin{equation*}
\Lambda(V,T,L)=\int_{0}^{\omega _{c}}\frac{d\omega }{\omega }[{f_{2}(\omega
+V)-f_{2}(\omega -V)}]\sin ^{2}{\omega L}\,,
\end{equation*}%
and it provides the linear response to the voltage $V$.
For $T=0$ we have $f_{2}(x)=|x|$ and the corresponding expressions %
\eqref{corrections} are given in the main text.

\section{\label{app:T} Extension to finite temperatures}

Consider now the case of finite $T$. The contributions to the
current at the lowest order in the interaction strength are still given by Eq.~(\ref{corrections}), but now
we put $\epsilon=0$ and the low-energy cutoff is provided by the function
$F(\omega,V)$ taken at finite $T$
\begin{equation*}
F(\omega ,V)=\frac{\omega +V}{\omega }\coth\frac{\omega +V}{2T}
-\frac{\omega - V}{\omega }\coth\frac{\omega -V}{2T}\,.
\end{equation*}%
The quantity $\mathcal{I}(\omega _{0},\epsilon ,V)$ then reads
 \begin{equation}
\begin{aligned}
\mathcal{I}(\omega _{0},T,V)& =\int_{0}^{\omega _{0}}d\omega F(\omega ,V)= \\
& =2V\left\{\ln \left( \tfrac{\omega _{0}}{2\pi T}\right) +1-\mbox{Re}\left[ \psi
\left( 1+\tfrac{iV}{2\pi T}\right) \right] \right\}  \\
& \simeq 2V \times \left\{
           \begin{array}{ll}
             \ln \dfrac{\omega _{0}}{2\pi T}, & T\gg |V|, \\[0.3cm]
             \ln \dfrac{\omega _{0} e}{V}, & T\ll |V|.
           \end{array}
         \right.
\end{aligned}
\label{finiteT}
\end{equation}
We are looking for scaling contributions to $\mathcal{I}(\omega _{0},T,V)$ as
functions of $T$. These can be written as
\begin{equation*}
\mathcal{I}(\omega _{0},T,V)\approx 2V\theta (T-c_{\ast }|V|)\ln \frac{\omega _{0}}{%
2\pi T} \,,
\end{equation*}%
where $c_{\ast}$ is a number of the order of unity.
We see that all the steps in the derivation for $T=0$ are
applicable to the case of finite temperature,
with the replacement $\epsilon \rightarrow T/c_{\ast}$.

\section{\label{app:unstable}  Unstable fixed points}

In this Appendix, we analyze the nonequilibrium scaling near the two unstable fixed points.
We first consider the solution of the RG equations near the fixed point $A$, where $G_a=1$ and $G_b=0$.
Linearizing in $\bar{G}_{a}=1-G_{a}\ll 1$,
we introduce
the combination $\bar{G}_{c}=\bar{G}_{a}-\frac{1}{4}G_{b}$  and  express
the RG equations as
 \begin{equation}
\begin{aligned}
\frac{d\bar{G}_{c}}{d\Lambda }& =2\alpha\bar{G}_{c}\theta (\epsilon -V_{a}) \,,  \\
\frac{dG_{b}}{d\Lambda }& =-\frac{\alpha_{3}}{2}G_{b}\theta _{+}(\epsilon )  \,.
\end{aligned}
\end{equation}
The solution to this set of equations reads:
\begin{equation}
\bar{G}_{c}(0)=\left[\bar{G}_{a}(\omega _{0})-\tfrac{1}{4}G_{b}(\omega_{0})\right](\omega _{0}/V_{a})^{2\alpha} \,,
\end{equation}%
and
\begin{equation}
G_{b}(0)=G_{b}(\omega _{0})(|V_{b}-V_a/2|/\omega _{0})^{\alpha_{3}} \, .
\end{equation}%
Notice that the scaling exponent of $G_{b}$ vanishes in the absence of interaction in the tip, $\alpha_{3} = 0$. 
It is nonzero in this case in the second order as $\alpha^{2}$, in agreement with Refs.~\cite{Aristov2010,gutman08}.
The conductance of the main wire is given by
\begin{equation}
G_{a}(0)=\bar{G}_{c}(0)+\tfrac{1}{4}G_{b}(0).
\end{equation}
We see that $\bar{G}_{c}(0)$ is increasing upon renormalization and thus the RG flow
leads away from the fixed point $A$, whereas $G_{b}\rightarrow 0$ in the limit $V_{a,b}\rightarrow 0$.

Another unstable fixed point is located at
$$G_{a}^{M}=\frac{\alpha+\alpha_{3}}{\alpha+2\alpha_{3}},\quad G_{b}^{M}=1-\left(\frac{\alpha}{\alpha+2\alpha_{3}}\right)^{2}.$$
The RG equations for
$\check{G}_{a}=G_{a}-G_{a}^{M}$ and $\check{G}_{b}=G_{b}-G_{b}^{M}$
have the form
\begin{multline}
\frac{d\check{G}_{a}}{d\Lambda } =\frac{\alpha}{2}
\left[\frac{4\alpha}{\alpha+2\alpha_{3}}%
\check{G}_{a}
+\check{G}_{b}\right]
\theta (\epsilon -V_{a}) \notag
\\
+\frac{\alpha_{3}}{2}%
\frac{\alpha+\alpha_{3}}{\alpha+2\alpha_{3}}\check{G}_{a}\theta _{+}(\epsilon )\,,
\\
\frac{d\check{G}_{b}}{d\Lambda } =\frac{\alpha_{3}(\alpha+\alpha_{3})}{(\alpha+2\alpha_{3})^{2}}%
\left[2\alpha\check{G}_{a}+(\alpha+2\alpha_{3})\check{G}_{b}\right]\theta _{+}(\epsilon )\,.
\end{multline}
All coefficients are positive, signaling a runaway flow of both $\check{G}_{a}$ and $\check{G}_{b}$.


\begin{thebibliography}{31}
\expandafter\ifx\csname natexlab\endcsname\relax\def\natexlab#1{#1}\fi
\expandafter\ifx\csname bibnamefont\endcsname\relax
  \def\bibnamefont#1{#1}\fi
\expandafter\ifx\csname bibfnamefont\endcsname\relax
  \def\bibfnamefont#1{#1}\fi
\expandafter\ifx\csname citenamefont\endcsname\relax
  \def\citenamefont#1{#1}\fi
\expandafter\ifx\csname url\endcsname\relax
  \def\url#1{\texttt{#1}}\fi
\expandafter\ifx\csname urlprefix\endcsname\relax\def\urlprefix{URL }\fi
\providecommand{\bibinfo}[2]{#2}
\providecommand{\eprint}[2][]{\url{#2}}

\bibitem[{\citenamefont{Tomonaga}(1950)}]{Tomonaga1950}
\bibinfo{author}{\bibfnamefont{S.}~\bibnamefont{Tomonaga}},
  \bibinfo{journal}{Progress of Theoretical Physics}
  \textbf{\bibinfo{volume}{5}}, \bibinfo{pages}{544} (\bibinfo{year}{1950}).

\bibitem[{\citenamefont{Luttinger}(1963)}]{Luttinger1963}
\bibinfo{author}{\bibfnamefont{J.~M.} \bibnamefont{Luttinger}},
  \bibinfo{journal}{Journal of Mathematical Physics}
  \textbf{\bibinfo{volume}{4}}, \bibinfo{pages}{1154} (\bibinfo{year}{1963}).

\bibitem{giamarchi04} T.~Giamarchi, {\it Quantum Physics in One Dimension}
(Oxford University Press, Oxford, 2004).

\bibitem[{\citenamefont{Kane and Fisher}(1992)}]{Kane1992}
\bibinfo{author}{\bibfnamefont{C.~L.} \bibnamefont{Kane}} \bibnamefont{and}
  \bibinfo{author}{\bibfnamefont{M.~P.~A.} \bibnamefont{Fisher}},
  \bibinfo{journal}{Phys. Rev. B} \textbf{\bibinfo{volume}{46}},
  \bibinfo{pages}{15233} (\bibinfo{year}{1992}).

\bibitem[{\citenamefont{Safi and Schulz}(1995)}]{Safi1995}
\bibinfo{author}{\bibfnamefont{I.}~\bibnamefont{Safi}} \bibnamefont{and}
  \bibinfo{author}{\bibfnamefont{H.~J.} \bibnamefont{Schulz}},
  \bibinfo{journal}{Phys. Rev. B} \textbf{\bibinfo{volume}{52}},
  \bibinfo{pages}{R17040} (\bibinfo{year}{1995}).

\bibitem[{\citenamefont{Furusaki and Nagaosa}(1996)}]{Furusaki1996}
\bibinfo{author}{\bibfnamefont{A.}~\bibnamefont{Furusaki}} \bibnamefont{and}
  \bibinfo{author}{\bibfnamefont{N.}~\bibnamefont{Nagaosa}},
  \bibinfo{journal}{Phys. Rev. B} \textbf{\bibinfo{volume}{54}},
  \bibinfo{pages}{R5239} (\bibinfo{year}{1996}).

\bibitem[{\citenamefont{Sassetti and Kramer}(1996)}]{Sassetti1996}
\bibinfo{author}{\bibfnamefont{M.}~\bibnamefont{Sassetti}} \bibnamefont{and}
  \bibinfo{author}{\bibfnamefont{B.}~\bibnamefont{Kramer}},
  \bibinfo{journal}{Phys. Rev. B} \textbf{\bibinfo{volume}{54}},
  \bibinfo{pages}{R5203} (\bibinfo{year}{1996}).

\bibitem[{\citenamefont{Egger et~al.}(2000)\citenamefont{Egger, Grabert,
  Koutouza, Saleur, and Siano}}]{Egger2000}
\bibinfo{author}{\bibfnamefont{R.}~\bibnamefont{Egger}},
  \bibinfo{author}{\bibfnamefont{H.}~\bibnamefont{Grabert}},
  \bibinfo{author}{\bibfnamefont{A.}~\bibnamefont{Koutouza}},
  \bibinfo{author}{\bibfnamefont{H.}~\bibnamefont{Saleur}}, \bibnamefont{and}
  \bibinfo{author}{\bibfnamefont{F.}~\bibnamefont{Siano}},
  \bibinfo{journal}{Phys. Rev. Lett.} \textbf{\bibinfo{volume}{84}},
  \bibinfo{pages}{3682} (\bibinfo{year}{2000}).

\bibitem[{\citenamefont{Dolcini et~al.}(2003)\citenamefont{Dolcini, Grabert,
  Safi, and Trauzettel}}]{Dolcini2003}
\bibinfo{author}{\bibfnamefont{F.}~\bibnamefont{Dolcini}},
  \bibinfo{author}{\bibfnamefont{H.}~\bibnamefont{Grabert}},
  \bibinfo{author}{\bibfnamefont{I.}~\bibnamefont{Safi}}, \bibnamefont{and}
  \bibinfo{author}{\bibfnamefont{B.}~\bibnamefont{Trauzettel}},
  \bibinfo{journal}{Phys. Rev. Lett.} \textbf{\bibinfo{volume}{91}},
  \bibinfo{pages}{266402} (\bibinfo{year}{2003}).

\bibitem[{\citenamefont{Dolcini et~al.}(2005)\citenamefont{Dolcini, Trauzettel,
  Safi, and Grabert}}]{Dolcini2005}
\bibinfo{author}{\bibfnamefont{F.}~\bibnamefont{Dolcini}},
  \bibinfo{author}{\bibfnamefont{B.}~\bibnamefont{Trauzettel}},
  \bibinfo{author}{\bibfnamefont{I.}~\bibnamefont{Safi}}, \bibnamefont{and}
  \bibinfo{author}{\bibfnamefont{H.}~\bibnamefont{Grabert}},
  \bibinfo{journal}{Phys. Rev. B} \textbf{\bibinfo{volume}{71}},
  \bibinfo{pages}{165309} (\bibinfo{year}{2005}).

\bibitem[{\citenamefont{Metzner et~al.}(2012)\citenamefont{Metzner, Salmhofer,
  Honerkamp, Meden, and Sch\"onhammer}}]{Metzner2012}
\bibinfo{author}{\bibfnamefont{W.}~\bibnamefont{Metzner}},
  \bibinfo{author}{\bibfnamefont{M.}~\bibnamefont{Salmhofer}},
  \bibinfo{author}{\bibfnamefont{C.}~\bibnamefont{Honerkamp}},
  \bibinfo{author}{\bibfnamefont{V.}~\bibnamefont{Meden}}, \bibnamefont{and}
  \bibinfo{author}{\bibfnamefont{K.}~\bibnamefont{Sch\"onhammer}},
  \bibinfo{journal}{Rev. Mod. Phys.} \textbf{\bibinfo{volume}{84}},
  \bibinfo{pages}{299} (\bibinfo{year}{2012}).
  
\bibitem{safi09} I. Safi, arXiv:0906.2363. 
  

\bibitem[{\citenamefont{Aristov et~al.}(2010)\citenamefont{Aristov, Dmitriev,
  Gornyi, Kachorovskii, Polyakov, and W\"olfle}}]{Aristov2010}
\bibinfo{author}{\bibfnamefont{D.~N.} \bibnamefont{Aristov}},
  \bibinfo{author}{\bibfnamefont{A.~P.} \bibnamefont{Dmitriev}},
  \bibinfo{author}{\bibfnamefont{I.~V.} \bibnamefont{Gornyi}},
  \bibinfo{author}{\bibfnamefont{V.~Y.} \bibnamefont{Kachorovskii}},
  \bibinfo{author}{\bibfnamefont{D.~G.} \bibnamefont{Polyakov}},
  \bibnamefont{and} \bibinfo{author}{\bibfnamefont{P.}~\bibnamefont{W\"olfle}},
  \bibinfo{journal}{Phys. Rev. Lett.} \textbf{\bibinfo{volume}{105}},
  \bibinfo{pages}{266404} (\bibinfo{year}{2010}).

\bibitem[{\citenamefont{Aristov}(2011)}]{Aristov2011}
\bibinfo{author}{\bibfnamefont{D.~N.} \bibnamefont{Aristov}},
  \bibinfo{journal}{Phys. Rev. B} \textbf{\bibinfo{volume}{83}},
  \bibinfo{pages}{115446} (\bibinfo{year}{2011}).

\bibitem[{\citenamefont{Aristov and W\"olfle}(2012)}]{Aristov2012}
\bibinfo{author}{\bibfnamefont{D.~N.} \bibnamefont{Aristov}} \bibnamefont{and}
  \bibinfo{author}{\bibfnamefont{P.}~\bibnamefont{W\"olfle}},
  \bibinfo{journal}{Lith. J. Phys.} \textbf{\bibinfo{volume}{52}},
  \bibinfo{pages}{89 } (\bibinfo{year}{2012}).

\bibitem[{\citenamefont{Aristov and W\"olfle}(2013)}]{Aristov2013}
\bibinfo{author}{\bibfnamefont{D.~N.} \bibnamefont{Aristov}} \bibnamefont{and}
  \bibinfo{author}{\bibfnamefont{P.}~\bibnamefont{W\"olfle}},
  \bibinfo{journal}{Phys. Rev. B} \textbf{\bibinfo{volume}{88}},
  \bibinfo{pages}{075131} (\bibinfo{year}{2013}).

\bibitem[{\citenamefont{Aristov and W\"olfle}(2014)}]{Aristov2014}
\bibinfo{author}{\bibfnamefont{D.~N.} \bibnamefont{Aristov}} \bibnamefont{and}
  \bibinfo{author}{\bibfnamefont{P.}~\bibnamefont{W\"olfle}},
  \bibinfo{journal}{Phys. Rev. B} \textbf{\bibinfo{volume}{90}},
  \bibinfo{pages}{245414} (\bibinfo{year}{2014}).

\bibitem[{\citenamefont{Yue et~al.}(1994)\citenamefont{Yue, Glazman, and
  Matveev}}]{Yue1994}
\bibinfo{author}{\bibfnamefont{D.}~\bibnamefont{Yue}},
  \bibinfo{author}{\bibfnamefont{L.~I.} \bibnamefont{Glazman}},
  \bibnamefont{and} \bibinfo{author}{\bibfnamefont{K.~A.}
  \bibnamefont{Matveev}}, \bibinfo{journal}{Phys. Rev. B}
  \textbf{\bibinfo{volume}{49}}, \bibinfo{pages}{1966} (\bibinfo{year}{1994}).

\bibitem[{\citenamefont{Maslov and Stone}(1995)}]{Maslov1995}
\bibinfo{author}{\bibfnamefont{D.~L.} \bibnamefont{Maslov}} \bibnamefont{and}
  \bibinfo{author}{\bibfnamefont{M.}~\bibnamefont{Stone}},
  \bibinfo{journal}{Phys. Rev. B} \textbf{\bibinfo{volume}{52}},
  \bibinfo{pages}{R5539} (\bibinfo{year}{1995}).

\bibitem[{\citenamefont{Oshikawa et~al.}(2006)\citenamefont{Oshikawa, Chamon,
  and Affleck}}]{Oshikawa2006}
\bibinfo{author}{\bibfnamefont{M.}~\bibnamefont{Oshikawa}},
  \bibinfo{author}{\bibfnamefont{C.}~\bibnamefont{Chamon}}, \bibnamefont{and}
  \bibinfo{author}{\bibfnamefont{I.}~\bibnamefont{Affleck}},
  \bibinfo{journal}{J. Stat. Mech.} \textbf{\bibinfo{volume}{2006}},
  \bibinfo{pages}{P02008} (\bibinfo{year}{2006}).
  
\bibitem{fnote1}  Some nonequilibrium properties of a Y-junction, in particular, 
the rectification of ac bias voltage, were studied within the bosonization
approach in C. Wang and D. E. Feldman, Phys. Rev. B 83, 045302 (2011).
  
\bibitem{nazarov03} Yu.V. Nazarov and L.I. Glazman, Phys. Rev. Lett. \textbf{91}, 126804 (2003)

\bibitem[{\citenamefont{Polyakov and Gornyi}(2003)}]{Polyakov2003}
\bibinfo{author}{\bibfnamefont{D.~G.} \bibnamefont{Polyakov}} \bibnamefont{and}
  \bibinfo{author}{\bibfnamefont{I.~V.} \bibnamefont{Gornyi}},
  \bibinfo{journal}{Phys. Rev. B} \textbf{\bibinfo{volume}{68}},
  \bibinfo{pages}{035421} (\bibinfo{year}{2003}).

\bibitem[{\citenamefont{Lal et~al.}(2002)\citenamefont{Lal, Rao, and
  Sen}}]{Lal2002}
\bibinfo{author}{\bibfnamefont{S.}~\bibnamefont{Lal}},
  \bibinfo{author}{\bibfnamefont{S.}~\bibnamefont{Rao}}, \bibnamefont{and}
  \bibinfo{author}{\bibfnamefont{D.}~\bibnamefont{Sen}},
  \bibinfo{journal}{Phys. Rev. B} \textbf{\bibinfo{volume}{66}},
  \bibinfo{pages}{165327} (\bibinfo{year}{2002}).

\bibitem{shi16} Zheng Shi and I. Affleck, Phys. Rev. B \textbf{94}, 035106 (2016);
Zheng Shi, J. Stat. Mech. \textbf{2016}, 063106 (2016). 

\bibitem[{\citenamefont{Aristov and W\"{o}lfle}(2008)}]{Aristov2008}
\bibinfo{author}{\bibfnamefont{D.~N.} \bibnamefont{Aristov}} \bibnamefont{and}
  \bibinfo{author}{\bibfnamefont{P.}~\bibnamefont{W\"{o}lfle}},
  \bibinfo{journal}{Europhysics Letters} \textbf{\bibinfo{volume}{82}},
  \bibinfo{pages}{27001} (\bibinfo{year}{2008}).

\bibitem[{\citenamefont{Aristov and W\"{o}lfle}(2009)}]{Aristov2009}
\bibinfo{author}{\bibfnamefont{D.~N.} \bibnamefont{Aristov}} \bibnamefont{and}
  \bibinfo{author}{\bibfnamefont{P.}~\bibnamefont{W\"{o}lfle}},
  \bibinfo{journal}{Phys. Rev. B} \textbf{\bibinfo{volume}{80}},
  \bibinfo{eid}{045109} (\bibinfo{year}{2009}).

\bibitem[{\citenamefont{Aristov and W\"olfle}(2011)}]{Aristov2011a}
\bibinfo{author}{\bibfnamefont{D.~N.} \bibnamefont{Aristov}} \bibnamefont{and}
  \bibinfo{author}{\bibfnamefont{P.}~\bibnamefont{W\"olfle}},
  \bibinfo{journal}{Phys. Rev. B} \textbf{\bibinfo{volume}{84}},
  \bibinfo{pages}{155426} (\bibinfo{year}{2011}).

\bibitem{aristov17}  D.~N. Aristov and P.~W\"olfle, to appear.  The renormalizability of the non-equilibrium model has been checked to second order, ${\cal O}(\alpha_{i}^{2})$. No two-loop contributions to the $\beta$-function have been found, similarly to the equilibrium case. \cite{Aristov2010}

\bibitem{gutman08} D.B. Gutman, Y. Gefen, and A.D. Mirlin, Phys. Rev. Lett. \textbf{101}, 126802 (2008);
Phys. Rev. B \textbf{80}, 045106 (2009). 

\end{thebibliography}
\end{document}